\documentclass[twoside]{ilcws10}
\usepackage[latin1]{inputenc}
\usepackage[dvips]{graphicx,epsfig,color}
\usepackage{wrapfig,rotating}
\usepackage{amssymb,amsmath,array}

\newcommand{\gsim}{\mbox{ \raisebox{-1.0ex}{$\stackrel{\textstyle >}
{\textstyle \sim}$ }}}
\newcommand{\lsim}{\mbox{ \raisebox{-1.0ex}{$\stackrel{\textstyle <}
{\textstyle \sim}$ }}}

\begin{document}
\title{
Probing the Majorana nature of TeV-scale radiative seesaw models at the 
ILC
}
\author{Mayumi Aoki$^1$ and Shinya Kanemura$^2$\footnote{Speaker}
\vspace{.3cm}\\
1- Institute for Theoretical Physics, Kanazawa University, 
Kanazawa 920-1192, Japan
\vspace{.1cm}\\
2- Department of Physics, University of Toyama, 
Toyama 930-8555, Japan\\
}

\maketitle

\begin{abstract}
 Two important features of TeV-scale radiative seesaw models, in
 which tiny neutrino masses are generated at the quantum level, 
 are an extended scalar (Higgs) sector and the Majorana nature.
 We study phenomenological aspects of these models at the ILC.
 It is found that
 the Majorana nature of the models can be tested directly
 via the electron-positron and electron-electron collision experiments.
\end{abstract}

\section{Introduction} 

Current neutrino data
give clear evidence for physics beyond the standard model (SM), 
indicating that neutrinos have tiny masses of the 0.1 eV scale.
Such tiny masses may be generated from dimension-five effective
operators 
   $\frac{c_{ij}}{2 \Lambda} \overline{\nu^c}_L^i {\nu}_L^j \phi^0
   \phi^0$, 
where $\Lambda$ is a mass scale, $c_{ij}$ are dimensionless
coefficients, and $\phi^0$ is the Higgs boson.
The neutrino mass matrix can then be given by 
   $M_\nu^{ij} = {c_{ij} \langle{\phi^0}\rangle^2}/{\Lambda}$. 
As the vacuum expectation value (VEV) $\langle \phi^0 \rangle$ of the Higgs
boson is ${\cal O}(100)$ GeV, the observed tiny neutrino masses
are realized when
$(c_{ij}/\Lambda) \sim {\cal O}(10^{-14})$ GeV$^{-1}$. 
How can we naturally explain such a small number? 
In the tree-level seesaw scenario, the Majorana masses of right-handed
neutrinos ($=\Lambda$) have to be set
much higher than the electroweak scale~\cite{see-saw}.

Quantum generation of neutrino masses can be an alternative 
to obtain $(c_{ij}/\Lambda)$ $\sim {\cal O}(10^{-14})$ GeV$^{-1}$.
Thanks to the loop suppression factor, $\Lambda$ in these models  
can be lower  as compared to that in the tree-level seesaw models.
Consequently, the tiny neutrino masses would be explained in a
natural way only by the TeV-scale dynamics.
The original model along this line was proposed by Zee~\cite{zee}, 
and some variations were
considered~\cite{radiative,ZeeBabu,knt,MaModel,aks-prl}, where
neutrino masses are induced at the one-, two- or three-loop level.
It must be  charming 
that they are directly testable at the LHC and the ILC.

 A general feature in radiative seesaw models
 is the extended Higgs sector. 
 Discovery of extra Higgs bosons and detailed measurements of
 their properties at future collider experiments can give partial 
 evidence for the models.  
 Previous works on phenomenology in radiative seesaw
 models\cite{Babu:2002uu,AristizabalSierra:2006gb,Nebot:2007bc}  
 mainly discuss the constraint on the flavor structure
 and collider physics for the Higgs sector~\cite{
 Deshpande:1977rw,CaoMa,LopezHonorez:2006gr,typeX,Belyaev:2009zd}.
 Next, the Majorana nature is another common feature.
 To induce tiny Majorana masses for neutrinos, 
 we need lepton number violating interactions in the Higgs
 sector~\cite{zee,ZeeBabu} or right-handed neutrinos with TeV-scale
 Majorana masses~\cite{knt,MaModel,aks-prl}.
 When the future data would indicate an extended Higgs sector
 predicted by a specific radiative seesaw model,
 direct detection of the Majorana property at colliders must be
 a fatal probe to identify the model.

 In this talk, we discuss phenomenology of TeV-scale radiative seesaw
 models, in particular a possibility of detecting the Majorana
 nature at collider experiments\cite{ak-letter}.
 We here discuss three typical radiative seesaw models as
 reference models; the model by Zee and Babu~\cite{ZeeBabu},
 that by Ma~\cite{MaModel}, and that in Ref.~\cite{aks-prl}.

\begin{figure}[t]
\includegraphics[width=14cm,angle=0]{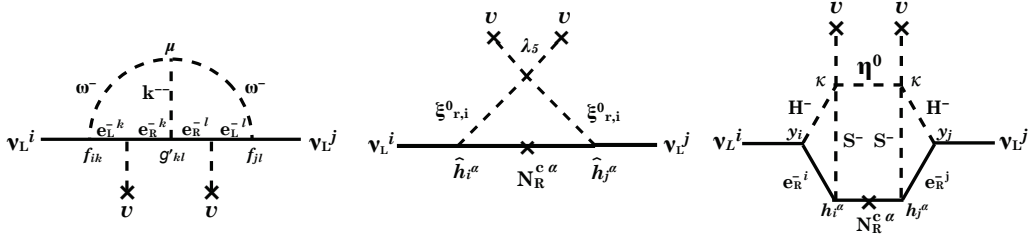}
  \caption{
 Feynman diagrams for neutrino masses in the model by Zee-Babu~\cite{ZeeBabu}
 (left), that by Ma~\cite{MaModel} (center) and that in Ref.~\cite{aks-prl} (right).
 }
  \label{fig:diag}
\end{figure}

\section{Radiative seesaw models}

\noindent
{\bf The Zee-Babu model} \hspace{2mm}

In the model proposed in Ref.~\cite{ZeeBabu} (we refer to as the
Zee-Babu model), in addition to singly-charged singlet scalar bosons
$\omega^\pm$, doubly-charged singlet scalar fields $k^{\pm\pm}$ are
introduced,  both of which carry
the lepton number of two unit.
%
The neutrino mass matrix is generated at the two-loop level via the 
diagram in Fig.~\ref{fig:diag} (left). 
The universal scale of neutrino masses is determined by the two-loop
suppression factor $1/(16\pi^2)^2$ and the lepton number violating
parameter $\mu$. The charged lepton Yukawa coupling
constants $y_{\ell_{i}}$ ($y_e \ll y_\mu \ll y_\tau \lsim 10^{-2}$)
give an additional suppression factor.
Thus, any of $f_{ij}$ or $g_{ij}$ in Fig.~\ref{fig:diag} (left) can be 
of ${\cal O}(1)$ when $m_\omega$ and $m_k$ are at the TeV scale.
The flavor structure of the mass matrix is determined
by the combination of the coupling constants $f_{ij}$ and $y_{i} g_{ij} y_{j}$.
The flavor off-diagonal coupling constants $f_{ij}$ and $g_{ij}$ induce lepton flavor violation (LFV).

In the scenario with hierarchical neutrino masses,
$f_{ij}$ satisfy $f_{e\mu} \simeq f_{e\tau} \simeq f_{\mu\tau}/2$.
The typical relative magnitudes among the coupling constants $g_{ij}$
can be
$g_{\mu\mu}: g_{\mu\tau}: 
g_{\tau\tau} \simeq 1 :  m_\mu/m_\tau: (m_\mu/m_\tau)^2$. 
For $g_{\mu\mu}\simeq 1$, the neutrino data and the LFV data give 
the constraints such as $m_k \gsim 770$ GeV and $m_\omega \gsim 160$
GeV~\cite{AristizabalSierra:2006gb}. 
On the other hand, the constraints on the
couplings and masses are more stringent for the inverted neutrino mass 
hierarchy. The current data then gives $m_\omega \simeq 825$ GeV
for $g_{\mu\mu}\simeq 1$~\cite{AristizabalSierra:2006gb}. 
One of the notable things in this case is the lower bound on
$\sin^22\theta_{13}$, which is predicted as around 0.002~\cite{Babu:2002uu}. 

\vspace{2mm}
\noindent
{\bf The Ma model} \hspace{2mm}

The model in Ref.~\cite{MaModel}, which we here refer to as the Ma model,
is the simplest radiative seesaw model with right-handed neutrinos
$N_R^\alpha$, in which the 
discrete $Z_2$ symmetry is introduced and its odd quantum number is
assigned to $N_R^\alpha$.
The Higgs sector is composed of two Higgs doublet fields, one of which 
($\Xi$) is $Z_2$ odd. As long as the $Z_2$ symmetry is exact, the neutral
components of $\Xi$ do not receive VEVs.
We have one SM-like Higgs boson $h$,
and four physical $Z_2$-odd scalar states; $\xi_r^{0}$ (CP-even), $\xi_i^{0}$ (CP-odd)
and $\xi^\pm$ as physical scalar states.
This $Z_2$ odd Higgs doublet is sometimes called as the inert Higgs doublet~\cite{Deshpande:1977rw}
or the dark scalar doublet~\cite{CaoMa}.
 The LEP II limits are studied in this model in Ref.~\cite{IDM-LEPII}.

The neutrino masses are generated at the one loop level via the diagram
depicted in Fig.~\ref{fig:diag} (center),
in which $Z_2$ odd particles, $\xi^0$ and $N_R^\alpha$, are in the loop.
The universal scale for neutrino masses is determined by the one-loop
suppression factor $1/(16\pi^2)$, the scalar coupling $\lambda_5$ and
the mass  $M_{N_R^{\alpha}}$ of the right-handed neutrinos.
 
In this model, there are two scenarios with respect to the DM candidate; i.e.,
the lightest right-handed neutrino $N_R^1$
or the lightest $Z_2$-odd neutral field ($\xi_r^{0}$ or $\xi_i^{0}$).
For both cases, there are parameter regions where the neutrino data are
adjustable without contradicting other phenomenological
constraints~\cite{Ma-DM}.
In this talk, we consider the case where the dark doublet
component $\xi_r^{0}$ is the DM
candidate. 
If the mass of the DM is around 50 GeV, 
the mass difference between $\xi_r^{0}$ and $\xi_i^{0}$ is given by 
about 10 GeV when $\lambda_5 \sim 10^{-2}$.  
In this case, $\hat{h}_{i}^\alpha \sim 10^{-5}$ ($i=e, \mu, \tau$) are required to satisfy
the neutrino data.
The relic abundance of such DM is consistent with the WMAP data~\cite{LopezHonorez:2006gr}.


\vspace{2mm}
\noindent
{\bf The AKS model} \hspace{2mm}

In the model in Ref.~\cite{aks-prl}, which we here refer to as the AKS
model, it is intended that not only
the tiny neutrino masses and DM but also baryon asymmetry
of Universe are explained at the TeV scale.
In addition to the TeV-scale right-handed neutrinos $N_R^\alpha$ ($\alpha=1,2$),
the Higgs sector is composed of $Z_2$-even two Higgs doublets $\Phi_i$
($i=1,2$), where the physical component fields are $H$ (CP-even), $A$ (CP-odd),
$H^\pm$ and $h$ (CP-even),  
and $Z_2$-odd charged singlets $S^\pm$ and a $Z_2$-odd neutral real singlet $\eta^0$.
The neutrino mass matrix is generated at the three-loop level via 
the diagram in Fig.~\ref{fig:diag} (right). 
The mass matrix has the three loop factor $1/(16\pi^2)^3$
with additional suppression by the charged lepton Yukawa couplings $y_i$. 
The electron associated couplings $h_e^{1,2}$ and
the scalar coupling $\kappa$ are of ${\cal O}(1)$ for $m_{N_R}^{1,2}\sim {\cal O}(1)$ TeV.
The Yukawa coupling constants $h_i^\alpha$ are hierarchical
as $h_e^{1,2} (\simeq {\cal O}(1)) \gg h_\mu^{1,2} \gg h_\tau^{1,2}$. 

The parameter sets which satisfy the current data from neutrino
oscillation, LFV, relic abundances of DM
and the condition for strongly first order electroweak phase transition 
are studied in Ref.~\cite{aks-prl}. 
To reproduce the neutrino data, the mass of $H^\pm$ should be 100 - 200 GeV.
This is an important prediction of the model. 
In order to avoid the constraint from $b\to s\gamma$, the Yukawa interaction
for the doublet fields takes the form of so-called 
Type-X~\cite{typeX} 
where only one of the doublets couples to leptons and the rest does to
quarks\cite{type-yukawa}.
The physics of the Type-X two Higgs doublet model (THDM) shows
many distinctive features from the other type of extended Higgs sectors.
For example, $H$ and $A$ decay mainly into $\tau^+\tau^-$ when
$\tan\beta \gsim 3$ and $\sin(\beta-\alpha)\simeq 1$~\cite{typeX}.
There are basically two DM candidates, $\eta^0$ and $N_R^\alpha$.
We here consider the case where $\eta^0$ is the DM\cite{aks-prl}.
Some characteristics of $\eta^0$ are discussed in Refs.~\cite{aks-cdms2}. 
The coupling constant of $S^+S^-h$ is required to be of ${\cal O}(1)$
for strongly first order phase transition,
whose indirect effect appears in the quantum correction to the $hhh$ coupling
constant as a large deviation from the SM prediction~\cite{aks-prl}.
As long as kinematically allowed, $S^\pm$ decays via $S^\pm \to H^\pm \eta^0$ by 100\%.

\section{Phenomenology in radiative seesaw models at the LHC}

The existence of the extra Higgs bosons such as
charged scalar bosons, which are a common feature of radiative
seesaw models, can be tested at the LHC.  
Details of the properties of such extra Higgs bosons
are strongly model dependent, so that we can distinguish
models via detailed measurements of extra Higgs bosons.
In addition, as the (SM-like) Higgs boson $h$ is expected to be
detected, its mass and decay properties are thoroughly measured~\cite{Aad:2009wy}.
The radiative seesaw models with a DM candidate
can also be indirectly tested via the invisible decay of $h$
as long as its branching ratio is more than about $25$\% for $m_h=$ 120 GeV
with ${\cal L}=$ 30 fb$^{-1}$~\cite{LHCinv}.
The phenomenological analyses at the LHC in each model are in the
literature~\cite{Babu:2002uu,AristizabalSierra:2006gb,Nebot:2007bc,
CaoMa,typeX,Belyaev:2009zd}.

At the LHC, via the physics of extra scalar bosons such as (singly and/or
doubly) charged Higgs bosons and CP-even Higgs bosons, the structure of
the extended Higgs sector can be clarified to some extent.
In addition, the invisible decay of the SM-like Higgs boson and 
the mass spectrum of the extra Higgs bosons 
would give important
indication for a possibility to a radiative seesaw scenario. 
However, although they would  be a strong indication of radiative seesaw models,
one cannot conclude that such Higgs sector is of the radiative seesaw
models. In order to further explore the possibility to such models,
we have to explore the other common
feature of radiative seesaw models, such as the Majorana nature. 
In the next section, we discuss a possibility of testing the Majorana
nature at ILC experiments.

\section{Phenomenology in radiative seesaw models at the ILC}

At the ILC, properties of the Higgs sector can be
measured with much better accuracy than at the LHC, so that
we would be able to reconstruct the Higgs potential in any
extended Higgs sector if kinematically accessible. 
Invisible decays of the Higgs boson can also be tested when the branching ratio
$B(h\to invisible)$ is larger than a few \%~\cite{Schumacher:2003ss}.
Furthermore, the Majorana nature in radiative seesaw models; {\it i.e.,}
the existence of TeV scale right-handed Majorana neutrinos or
that of LFV interaction,
would also be tested.

\subsection{Electron-positron collisions}

In the pair production of charged scalar bosons at the $e^+e^-$
collision, which appear in the radiative seesaw models ($\omega^+\omega^-$
 in the Zee-Babu model, $\xi^+\xi^-$ in the Ma model, and $S^+S^-$ (and
 $H^+H^-$) in the AKS model), there are diagrams
of the $t$-channel exchange of left-handed neutrinos
or right-handed neutrinos
in addition to the usual Drell-Yan type $s$-channel diagrams.
The contribution of these $t$-channel diagrams is one of the discriminative
features of radiative seesaw models, and no such contribution enters into
the other extended Higgs models such as the THDM.
These $t$-channel effects show specific dependences on
the center-of-mass energy $\sqrt{s}$ in proportion to $\log s$
in the production cross section, and enhances the production
rates of the signal events for higher values of $\sqrt{s}$.
The final states of produced charged scalar boson pairs
are quite model dependent but with missing energy; 
\begin{eqnarray}
&&  e^+e^- \to \omega^+\omega^- 
  \to \ell_L^+\ell_L^- \underline{\overline{\nu}_{L} \nu_L^{}}\, ,
  \hspace{52mm} [{\rm The~Zee}{\rm-Babu~model}]\nonumber\\ 
&&  e^+e^- \to \xi^+\xi^- \to W^{+(\ast)}W^{-(\ast)} \xi_r^0 \xi_r^0 \to
 jjjj(jj\ell_L\underline{\nu_L}) \underline{\xi_r^0 \xi_r^0}\, , 
  \hspace{22mm} [{\rm The~Ma~model}]\nonumber\\
&&  e^+e^- \to S^+S^- \to H^+H^-\eta^0\eta^0 \to \tau^+_R\tau^-_R\underline{\overline{\nu}_L \nu_L \eta^0\eta^0}\, , 
 \hspace{30mm}  [{\rm The~AKS~model}] \nonumber
 \end{eqnarray}
where underlined parts in the final states are observed as missing energy.

\begin{figure}[t]
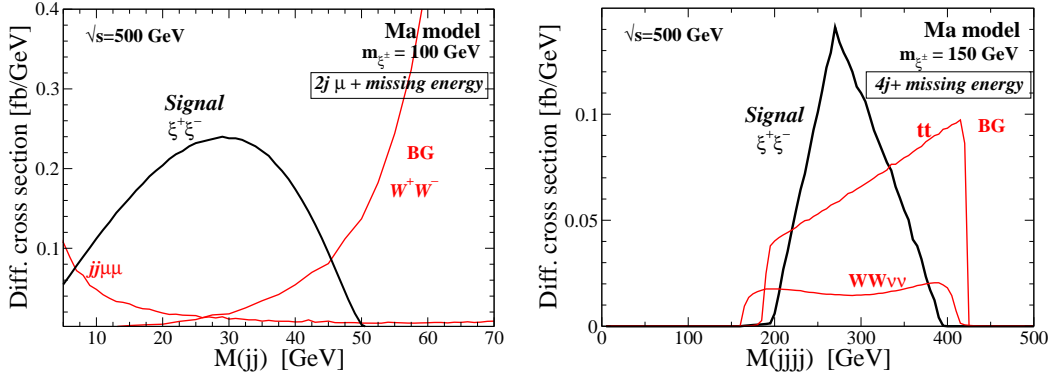

\includegraphics[width=6.6cm,angle=0]{Mjj.eps}
\hspace{2mm}
\includegraphics[width=6.8cm,angle=0]{M.eps}
  \caption{
The jets invariant mass distributions of the production rates of the signal in the Ma model at $\sqrt{s}=500$ GeV. 
{\it left} : The di-jet invariant mass $M(jj)$  distribution of the signal $e^+e^-\to \xi^+\xi^-  \to jj\mu\nu\xi^0_r\xi_r^0$ for $m_{\xi^\pm}=100$ GeV. {\it right} : $M(jjjj)$ distribution of $e^+e^- \to \xi^+\xi^- \to W^+W^- \xi_r^0 \xi_r^0 \to jjjj \xi_r^0$ for $m_{\xi^\pm}=150$ GeV. In addition to the rate from the signal process, those for main backgrounds are also shown.
 }
  \label{fig:ma}
\end{figure}

\vspace{5mm}
\noindent
\underline{\it The Ma model}: 
The coupling constants $\hat{h}_{e}^\alpha$ ($\alpha=1,2$)
are strongly constrained from neutrino data and LFV data. 
As a typical choice of parameters, we consider
$m_{\xi_r}=50$ GeV,  $m_{\xi_i}=60$ GeV,
$m_{\xi^\pm} \sim 100$ GeV, 
$m_{N_R^1}  = m_{N_R^2}  
 = 3 {\rm ~TeV}$, 
$\lambda_5 = -1.8 \times 10^{-2}$,
$\hat{h}_e^\alpha, \hat{h}_\mu^\alpha, \hat{h}_\tau^\alpha
\sim 10^{-5}$, 
in which the normal neutrino mass hierarchy is realized.
Because $\hat{h}_e^\alpha$ are very small for a TeV scale $m_{N_R^\alpha}$,
the contribution of the $t$-channel diagrams to the signal $e^+e^-\to\xi^+\xi^-$ 
is much smaller than that from Drell-Yan type diagrams.
For most of the possible values of $\hat{h}_\ell^\alpha$ and $m_{N_R^\alpha}^{}$
  which satisfy the LFV and the neutrino data, the contribution of the
  $t$-channel diagrams is negligible.
The production cross section of a charged Higgs pair $\xi^+\xi^-$
is therefore similar to that in the usual THDM:
about 92 (10) fb for $m_{\xi^\pm}=100$ (150) GeV at $\sqrt{s}=$ 500 GeV.
The produced $\xi^\pm$ decay into $W^{\pm (*)} \xi_{r,i}^0$.

 In Fig.~2 ({\it left}), we show the invariant
mass distribution of the di-jet $jj$ of the production cross section of the
signal, $e^+e^-\to \xi^+\xi^- \to W^{+ \ast} W^{-\ast} \xi_r^0 \xi_r^0 \to jj\mu\nu\xi^0_r\xi_r^0$
for $m_{\xi^\pm}=100$ GeV.
The main backgrounds come from $WW$.
The $jj\mu\mu$ events from $ZZ$, $\gamma\gamma$, and $Z\gamma$
can also be the backgrounds.
A factor of 0.1 is multiplied to the rate of the $jj\mu\mu$ backgrounds
for the miss-identification probability of a muon. 
The signal is significant around $M(jj)\sim$30 GeV. 
The invariant mass cut (such as 15 GeV$< M(jj) <$ 40 GeV) is effective 
to reduce the backgrounds.
For the numerical evaluation, we have used a package CalcHEP 2.5.4~\cite{calchep}.

For $m_{\xi^\pm} > m_W+m_{\xi_r}$, on the other hand, 
the signal $W^+W^-\xi^0_r\xi^0_r$ can be measured by
detecting the events of four jets with a missing energy. The main background
comes from $W^+W^-\nu\nu$ and $t\overline{t}$. 
By the invariant
mass cuts of two-jet pairs at the $W$ boson mass, the biggest background
from $WW$ can be eliminated. In Fig.~2 ({\it right}), we show the invariant
mass distribution of $jjjj$ of the production cross sections of the
signal and the backgrounds without any cut.
A factor of 0.1 is multiplied to the rate of $tt$ background, by which
the probability of the lepton from a $W$ that escapes from detection is
approximately taken into account.  
The signal is already significant. The invariant mass cut
($M(jjjj)< 300$ GeV) gives an improvement for the signal/background ratio.

\begin{figure}[t]
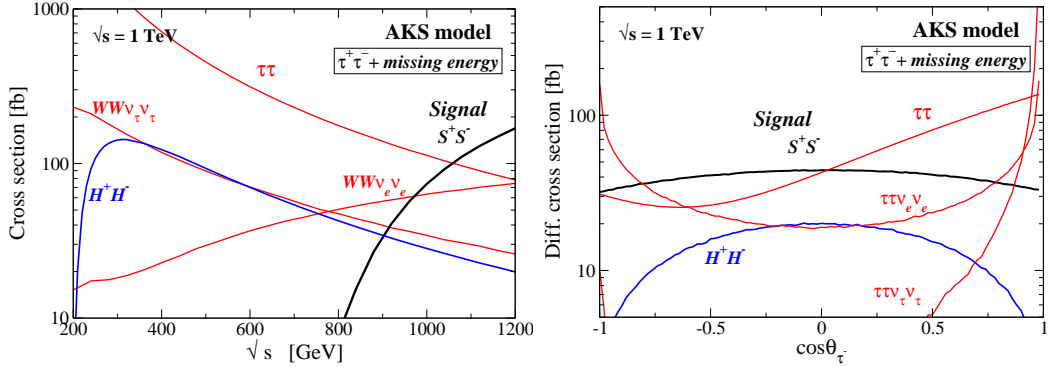

\includegraphics[width=7cm,angle=0]{roots.eps}
\includegraphics[width=6.7cm,angle=0]{AKS_angle.eps}
  \caption{
{\it left} : The cross sections of the signal, $e^+e^-\to S^+S^- \to
\tau^+\tau^-$ (+ missing energy), in the AKS model as a function of 
the collision energy $\sqrt{s}$.
{\it right} : The differential cross section of the signal for $\sqrt{s}=$ 1 TeV as a function of the
angle of the direction of the outgoing $\tau^-$ and the beam axis of incident electrons.
 In addition to the rate from the signal, those from backgrounds 
 such as $\tau^+\tau^-$, $\tau^+\tau^- \overline{\nu} \nu$ and
 $H^+H^-$ are also shown.
 }
  \label{fig:aks}
\end{figure}

\vspace{5mm}
\noindent
\underline{\it The AKS model}: 
We take a typical successful scenario for the neutrino data
with the the normal mass hierarchy, 
the LFV data and the DM data as well as the condition for strongly first order phase
transition~\cite{aks-prl}; 
$m_{\eta}=50$ GeV, $m_{H^\pm}=100$ GeV, $m_{S^\pm}=400$ GeV, 
$m_{N_R^1}^{}=m_{N_R^2}^{}=3$ TeV, 
$\kappa \sim{\cal O}(1)$, 
$h_{e}^1=h_{e}^2=2 \gg h_{\mu}^1$, $h_{\mu}^2  \gg h_{\tau}^1$, $h_{\tau}^2$, 
$\sin(\beta -\alpha)=1$, $\tan\beta=10$.
%
Because $h_e^{1,2}\sim$ ${\cal O}$(1), the contribution
from the $t$-channel $N_R^\alpha$ exchange diagrams
to the production cross section of $S^+S^-$ dominate that from the 
Drell-Yan diagrams~\cite{aks-prl}. The cross section is
about 87 fb for $m_{S^\pm} = 400$ GeV at $\sqrt{s}=1$ TeV. 
As the decay branching ratio of $S^\pm \to H^\pm \eta^0$ is 100\% and
that of $H^\pm \to \tau^\pm \nu$ is also almost 100\% because of the
Type-X THDM interaction for $\tan\beta = 10$, the final state of the
signal is $\tau^+\tau^- \overline{\nu} \nu \eta^0\eta^0$ with almost the same
 rate as the parent $S^+S^-$ production.
The main SM backgrounds are
$\tau^+\tau^-$ and $\tau^+\tau^- \overline{\nu}\nu$.
The pair production of the doublet like charged Higgs boson $H^+H^-$
can also be the background. 
As the signal rate dominantly comes from the $t$-channel diagram,
it becomes larger for larger $\sqrt{s}$, while
the main backgrounds except for $\tau\tau\nu_e\nu_e$
are smaller because they are dominantly $s$-channel processes (Fig.~3 ({\it left})). 
At $\sqrt{s}=1$ TeV, the rate of the signal without cut
is already large enough as compared to those of the backgrounds.
It is expected that making appropriate kinematic cuts  will
improve the signal background ratio to a considerable extent.
The $\sqrt{s}$ scan will help us to confirm that the signal rate
comes from the $t$-channel diagrams. Fig.~3 ({\it right}) shows the differential cross 
section of the signal at $\sqrt{s}=$ 1 TeV as a function of $\cos\theta_{\tau^-}$, where
$\theta_{\tau^-}$ is the angle between the direction of the outgoing $\tau^-$ and 
the beam axis of incident electrons.
The distribution of the background from $\tau\tau$ is 
asymmetric, so that the angle cut for larger $\cos\theta_{\tau^-}$ reduces the backgrounds.

\subsection{Electron-electron collisions}

The ILC has a further advantage to test radiative seesaw models via the
experiment at the $e^-e^-$ collision option, where dimension five
operator of $e^-e^- \phi^+\phi^+$, which is 
the sub-diagram of the 
loop diagrams for neutrino mass matrix. 
This direct test of the dimension five operator is essential
to identify the radiative seesaw models.

The Majorana nature in the Zee-Babu model is in the lepton number violating
coupling constant $\mu$ of $k^{++} \omega^-\omega^-$, which generates the dimension five
operator of $e^-e^- \omega^+ \omega^+$ at the tree level
via the $s$-channel $k^{--}$ exchange.
The cross section of $e^-e^- \to \omega^-\omega^-$ is given by
\begin{eqnarray}
 \sigma(e^-e^- \to \omega^-\omega^-) = \frac{1}{8 \pi}
  \sqrt{1-\frac{4 m_{\omega}^2}{s}} 
  \frac{\mu^2 g_{ee}^2}{(s-m_{k}^2)^2+m_{k}^2 \Gamma_{k}^2},
 \end{eqnarray}
where $\Gamma_{k}$ is the total width of $k^{\pm\pm}$.  
In the Ma model and the AKS model the operator
comes from the $t$-channel $N_R^\alpha$ exchange diagram.
The cross section is evaluated as
\begin{eqnarray}
  \sigma(e^-e^- \to \phi^-\phi^-) =
      \int_{t_{\rm min}}^{t_{\rm max}} dt \frac{1}{128\pi s}
           \left|\sum_{\alpha=1}^n  (|c^\alpha|^2
           m_{N_R^\alpha}) \left(\frac{1}{t-m_{N_R^\alpha}^2}+\frac{1}{u-m_{N_R^\alpha}^2}\right) \right|^2,
 \end{eqnarray}
 where $n$ is the number of generation of right-handed neutrinos,
 $\phi^-$ represents the $Z_2$-odd charged scalar boson
 $\xi^-$ in the Ma model and $S^-$ in the AKS model. The constants
 $c^\alpha$ represent $\hat{h}_e^\alpha$ or $h_e^\alpha$ in the Ma
 model or the AKS model, respectively.
Due to the Majorana nature of the $t$-channel diagram, we
 obtain much larger cross section in the $e^-e^-$ collision than in the
 $e^+e^-$ collision in each model. 

\begin{wrapfigure}{r}{0.48\columnwidth}
\centerline{\includegraphics[width=0.45\columnwidth]{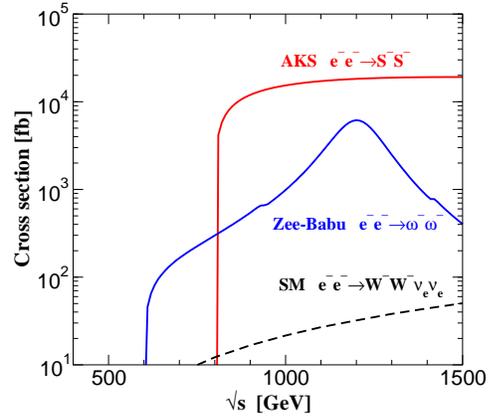}}
   \caption{Cross sections of like-sign charged Higgs pair
 productions in the Zee-Babu model and in the AKS model. 
 } 
   \label{fig:ee}
 \vspace{-3mm}
\end{wrapfigure}
The mass matrix of left-handed neutrinos is generated at the one, two and three loop levels
in the Ma model, the Zee-Babu model and the AKS model, respectively.
Therefore, the coupling constants can be basically hierarchical among the models, so are
the cross sections.
For the typical scenarios in these models,
the cross sections are shown in Fig.~4. 
The rate of $\omega^-\omega^-$ in the Zee-Babu model can be larger than
several times 100 fb for 800~GeV $\lsim \sqrt{s} \lsim 1.5$~TeV, 
where $m_{\kappa}=1200$~GeV, $\mu=800$ GeV and $g_{ee}=0.17$ are asuumed, 
and  $\Gamma_k$ is computed as about 168~GeV 
in our scenario (see \cite{ak-letter}).
It becomes maximal (several times pb) at $\sqrt{s}\sim m_k$,
and above that asymptotically reduces by $1/s$.
The maximal value of the cross section is sensitive to the value of
$g_{ee}$ and $\mu$. 
The signal should be like-sign dilepton with a missing energy.
On the other hand, in the Ma model, production cross sections of $e^-e^- \to
\xi^-\xi^-$ are smaller than $10^{-3}$ fb.
Hence, most of the successful scenarios in the Ma model
the process $e^-e^- \to \xi^-\xi^-$ is difficult to be seen. 
In the AKS model, the cross section of $e^-e^-\to S^-S^-$ 
is large, and its value amounts to about 15 pb at $\sqrt{s}=1$ TeV in the
present scenario.
Above the threshold, the magnitude of
the cross sections are not sensitive to $\sqrt{s}$,
so that even if $m_{S^\pm}^{}$ would be at the TeV scale, 
we might be able to test it
at future multi-TeV linear colliders, such as the Compact Linear Collider~\cite{CLIC}.
Because $B(S^\pm \to \eta^0 H^\pm) \simeq B(H^\pm \to \tau^\pm \nu) \simeq 100$~\%,
the signal should be $\tau^-\tau^-\nu\nu\eta\eta$ with almost the same
rate as long as $m_{S^\pm}^{} < m_{N_R^\alpha}$.  
The background mainly comes from $W^-W^-\nu_e\nu_e$, 
and the cross section is about 2.3 fb (22 fb) for $\sqrt{s}=500$
GeV (1 TeV). The branching ratio for the leptonic decay of $W$ bosons is 30\%, so that the
rate of the final state $\ell\ell'\nu\nu\nu\nu$ is at most 2 fb.
Therefore, the signal in the AKS model and in the Zee-Babu model can be
seen.

There are several other models with lepton number violating
interactions or right-handed Majorana neutrinos.
Atwood et al. have discussed the signature of heavy Majorana neutrinos
in the model without $Z_2$ symmetry via charged Higgs pair production at
$e^+e^-$ and $e^-e^-$ collisions~\cite{atwood}.
In supersymmetric models, Majorana particles also appear, and their
effects also give similar $t$-channel contributions 
in the slepton pair production $e^-e^- \to \tilde{e}^- \tilde{e}^-$, 
whose cross section is at most ${\cal O}$(100) fb.
The final state would be $e^-e^- \chi^0\chi^0$ for example.

\section{Conclusion}

We have discussed 
general features of TeV-scale radiative seesaw models. They are 
characterized by an extended scalar (Higgs) sector and the Majorana nature.
Various phenomenological aspects especially in experiments at the ILC.
have been discussed in the Zee-Babu model (2-loop), the Ma model (1-loop),
and the 3-loop model in Ref.~\cite{aks-prl}.
While the extended Higgs sector can be explored at the LHC, 
the Majorana nature of the models can directly be tested at the ILC
via the pair production of the charged scalar bosons at the
electron-positron and electron-electron collision experiments.


\begin{footnotesize}


\end{footnotesize}


\end{document}